# WikiTermBase: An AI-Augmented Term Base to Standardize Arabic Translation on Wikipedia

**Michel Bakni**
ESTIA, Institute of Technology
m.bakni@estia.fr

**Abbad Diraneyya**
Columbia University
amd2363@columbia.edu

**Wael Tellat**
Independent freelancer
wael.tellat@gmail.com

## Abstract

Term bases are recognized as one of the most effective components of translation software in time saving and consistency. In spite of the many recent advances in natural language processing (NLP) and large language models (LLMs), major translation platforms have yet to take advantage of these tools to improve their term bases and support scalable content for underrepresented languages, which often struggle with localizing technical terminology. Language academies in the Arab World, for example, have struggled since the 1940s to unify the way new scientific terms enter the Arabic language at scale. This abstract introduces an open source tool, WikiTermBase, with a systematic approach for building a lexicographical database with over 900K terms, which were collected and mapped from a multitude of sources on semantic and morphological basis. The tool was successfully implemented on Arabic Wikipedia to standardize translated English and French terms.

## Introduction

With 70 million articles in over 300 languages, Wikipedia is one the most multilingual knowledge repositories ever made (Samoilenko, 2021). In February 2025, Arabic Wikipedia had over 750,000 paragraphs in 85,000 articles translated from English alone, according to the Content Translation statistics (Wikimedia Foundation, 2025), a tool with which a fraction of Wikipedia's vast multilingual content is created.

So far, no Wikipedia edition has a formal term base. Wikipedia's content is the result of massive crowdsourced efforts by about 100,000 unique volunteers (Ozurumba, 2022). Individual editors each use their own lexicographic resources, often spending extensive amounts of time (~20% of content creation time, according to our research) looking up terms that have no standard equivalents. This leads to a lack of consistency in translated content, and a lack of knowledge sharing among translators, both of which are typically solved by term bases (Lewis & O'Connor, 2012).

Inconsistent terminology has long been recognized as a leading problem for multilingual content with a significant impact on quality. In Arabic, the lack of terminology standards has been reported as a serious hindrance to readability, rendering translations "not only useless but misleading as well." (Saraireh, 2001). Zarzar (2017) cites ten common Arabic translations of the English term "brake", only for its sense as a "vehicle's brake". Studies have found that this lack of consistency is a huge problem for translators (Merkel, 1996).

Many term bank projects attempted to address this problem. In an extensive study, Matuschek et al (2013) proposed the following requirements for a strong term base: 1. It can be continuously expanded, 2. It has a variety of lexicographic information, and 3. It is seamlessly integrated into a translation interface. Most existing tools fail this test. Significant past attempts include Arabterm by ALECSO, and the EuroTermBank Toolkit, a successful database of 14.5 million terms (Lagzdiņš et al, 2022). Wikidata, a Wikipedia sister project, has a lexicographical data extension, but it still suffers from limitations, including multi-component phrases (Morshed, 2021). These experiences, however, severely lack the integration of advanced NLP transformers and cloud technologies, thus heavily leaning on manual labor, which renders the content expensive and the scalability poor. For example, Wikidata still has only 10K lexemes for English over six years after launching its lexicographic data extension (Wikidata, 2025).

## Methods

This proposal introduces a term base that enables translators to make informed decisions by consulting a vast collection of terms, ranked according to morphological frequency and semantic similarity. The data was extracted from about 50 digitized dictionaries with over 950,000 entries in Arabic, English and French. Alongside a recommended equivalent, the retrieved results include definitions, lexicographical details, and bibliographical information.

Users can query the term base directly from within Wikipedia's editing interface. The term base is accessed through SQL queries sent to a MariaDB instance hosted on Toolforge, a Wikimedia cloud service. In the future, the term base may be migrated to Wikidata to integrate it into a collaborative wiki project. This, however, is contingent on the feasibility of transferring our massive data set into the Lexicographic Data extension, and on the community's willingness to deal with a data dump of such scale.

To enable standardizations, the term base groups terms morphologically and semantically. First, the user's query is used to fetch all matching source language terms. The Arabic equivalents of each of those terms, already mapped in the database, are then grouped by morphological form, after clean up. The user is recommended the most frequent form across all dictionaries. In some cases, over 20 sources will agree on the standard equivalent.

LLMs, including Claude Sonnet 3.7 and GPT-4, were used for semantic grouping. This experimental approach has the advantage of ranking Arabic equivalents by contextual meaning, rather than only morphological similarity, resulting in more accurate standards. We used Wiktionary as a reference dictionary, from which we extracted a list of senses for each English term, assigned each a unique ID, and mapped it into Arabic equivalents. A term like "scale" has at least five unique senses in our term base, some of which share the same Arabic equivalent. of the 950K terms in the database, at least 20% are considered duplicates

Wiktionary has been chosen as the anchor for three reasons. First, it is among the most comprehensive dictionaries available, with over 1.3 million entries for English alone (Wiktionary, 2025). Second, Wiktionary sense definitions are far more granular than any other dictionary we surveyed, for example: the word "absorb" has 4 senses listed in Cambridge, 9 in Merriam-Webster, and 14 in Wiktionary. Third, it is an open wikiproject with close ties to Wikipedia.

The query processing includes the following detailed steps (see Table 1 for an actual output example):

1. **Query the database**. Search the source language term using a SQL query. The query will return the exact match of the term (if available), and all similar entries, ranked by edit distance.
2. **Retrieve foreign terms**. Group the results for each unique source language term. The database assigns an ID for each morphologically unique term.
3. **Rank by frequency**. Rank the senses of each source language term by their frequency in the database. Each unique term has a subsequence of IDs for its predefined senses, extracted from our reference lexicographical database, Wiktionary.
4. **Map semantic equivalents**. Retrieve the Arabic equivalents mapped to each sense ID. Arabic term definitions were mapped in advance to sense IDs from multilingual dictionaries using LLMs.
5. **Output the results**. Rank the Arabic equivalents for each sense based on a statistical analysis. This ranking, again, groups morphologically identical terms, after preprocessing to remove diacritics, special characters, and foreign language characters.

## Results

The proposed term base enables Arabic Wikipedia editors to find standard equivalents of technical terms, and to cite them using reliable lexicographical resources. Table 1 shows a detailed sample of the output data following the processing method proposed in Figure 1. The tool is open source, with the code available on a Github repository and the functions integrated into a Wikipedia add-on.[1]

Before building the tool, we conducted initial user research. Interviews with Wikipedia editors and Arabic linguists directly recommended a term base solution, estimating that 20 - 30% of translation time was spent manually looking up or devising translations for technical terms. The requirements collected from this Interviews guided the tool's development.

After the public release in February 2025 (Figure 2), user testing yielded tremendously positive feedback. With a sample representing 20% of the tools's current user base (80 editors), the tool received 90% positive ratings for making translation quicker, easier, and more standard (Figure 3). The main areas of improvement are multi-word search, and in opening up the database for user contributions. Users estimated that the tool saved them about 20% of the time needed to translate a one page (250 word) long article. Less than a month after launch, the tool has been referenced in organic community discussions of term standardization.[2]

Currently, we are going through a multi-step evaluation of the LLM accuracy for semantic mapping. Tentative results are promising, with potential improvements in domain tagging, and expanding mapping into derived terms.

## Conclusions

This paper introduced a tool to standardize Arabic terms based on search queries in a relational database compiled from extensive lexicographical resources, and augmented with LLMs for semantic tagging. The tool has already been tested with nearly 100 Arabic Wikipedia users, who reported improvements in their translation experience, and resulted in much higher consistency in terminology choices.

A future improvement is adding AI-powered semantic processing (in step 1) and morphosyntactic processing (in step 4). In practice, this would involve taking the embedding of an entire input sentence with a source language term, mapping it to a granular sense, then exclusively returning a target language equivalent for that sense. Due to the morphosyntactic complexity of Arabic, additional processing can also add useful information about synonyms and the Arabic term's structure, helping automatically format the output, therefore completing the end-to-end translation and ensuring full consistency in future content.

---

[1] Github repo: https://github.com/forzagreen/wikitermbase

[2] See: https://ar.wikipedia.org/wiki/نقاش:حساسات_الروبوتات

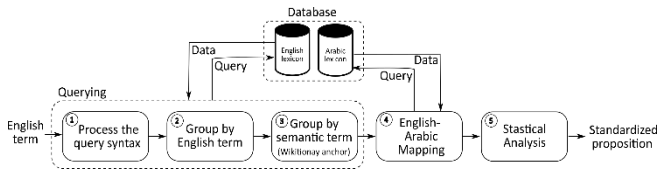

Figure 1: Method of the research

| Step ID | Output |
|---|---|
| 1 | **Adsorption**<br>Carbon adsorption<br>Adsorption drying<br>Adsorption medium etc. …. |
| 2 | **Adsorption: 25 instances**<br>Carbon adsorption: 5 instances<br>Adsorption drying: 2 instances…. |
| 3 | **Adsorption (physics): 15 instances**<br>Adsorption (chemistry): 7 instances<br>Adsorption (other): 3 instances… |
| 4 | Adsorption (physics): 15 instances<br>**12 instances:** امتزاز "Imtzāz" /ʔmtzaːz/<br>2 instances: ادمصاص "Idmṣāṣ" /ʔdmsˤaːsˤ/<br>1 instances: تكثيف "Takthīf" //... |
| 5 | امتزاز |

Table 1: Sample of results.

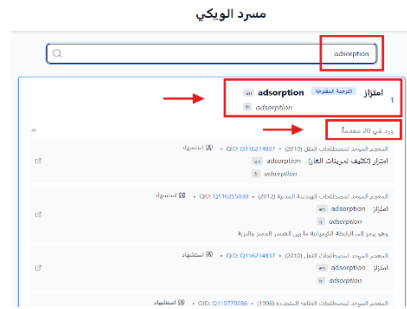

Figure 2. The tool's interface on Wikimedia's Toolforge, which is also accessible directly through Arabic Wikipedia

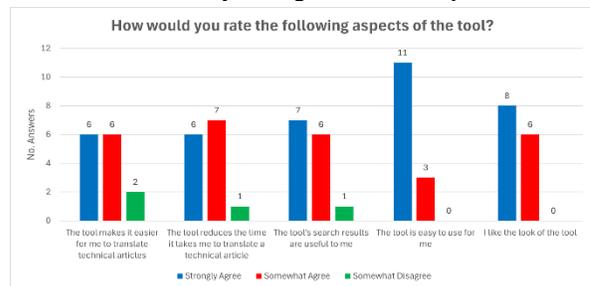

Figure 3. User evaluation results in a May 2025 survey, with 15 participants.